# Minimal Immunogenic Epitopes Have Nine Amino Acids

J. C. Phillips

Dept. of Physics and Astronomy, Rutgers University, Piscataway, N. J., 08854

Abstract

**Background**

To be cost-effective, biomedical proteins must be optimized with regard to many factors. Road maps are customary for large-scale projects, and here descriptive methods based on bioinformatic fractal thermodynamic scales are tested against an important example, HPV vaccine.

**Results**

Older scales from before 2000 are found to yield inconclusive results, but modern bioinformatic scales are amazingly accurate, with a high level of internal consistency, and little ambiguity. The dramatic effect of the single amino acid mutation D202H on stable HPV vaccine effectiveness is a long-standing mystery, which profiling with fractal $\Psi$ (aa,W) solves.

**Conclusions**

Most biomedical differences are too small to be resolved with any $\Psi$ scale other than the best bioinformatic scale. Bioinformatic scaling is also supportive of emergent microarray epitopic blood profiling for early cancer detection.

## Background

It is widely believed that cancerous mutations are formed initially as a few isolated cells with mutated DNA whose growth into tumors is controlled by autoantibodies. There are around ten million known cancer mutations, in principle controlled by a similar number (or more) of possible antibodies. Monoclonal antibodies with anti-cancer and anti-inflammatory functions are expensive, and recently this has turned attention to simplifying antibody-antigen interactions by focusing on selected antigenic proteins instead. Small antigenic regions (epitopes) interact with small antibody regions (paratopes). Small peptide epitope sequences are printed cost-effectively on microarrays. Due to their miniature format they allow for the multiplex analysis of several thousands of peptides at the same time while requiring a minimal sample volume [1].

We have identified epitopic features in scans of several well studied proteins to universal hydropathic properties, as quantified very accurately by a modern thermodynamic scale Ψ. Protein globular shapes are determined by competing hydrophobic forces (pushing phobic segments towards the globular cores) and hydrophilic forces (pushing philic segments towards the globular water interface). Moreover, the leading physicochemical properties determining protein aggregative mutations are hydrophobicity, secondary structure propensity and charge [2]. To quantify these effects in the classic period of molecular biology (before 2000), no less than 127 hydropathicity scales were proposed, but seldom compared for accuracy and applicability [3].

## Methods

The modern hydropathicity scale Ψ, built by Brazilian bioinformaticists Moret and Zebende (MZ), is an interdisciplinary bridge connecting proteins to statistical mechanics and phase diagram critical points [4]. They evaluated solvent-exposed surface areas (SASA) of amino acids in > 5000 high-resolution (< 2A) protein segments, and fixed their attention on the central amino acid in each segment. The lengths of the small segments L = 2N + 1 varied from 3 to 45, but the interesting range turned out to be $9 \leq L \leq 35$. Across this range they found linear behavior on a log-log plot for each of the 20 amino acids (aa):

$$\text{logSASA(L)} \sim \text{const} - \Psi(aa) \, \text{const.logL} \qquad ( \, 9 \leq L \leq 35) \qquad (1)$$

Here Ψ(aa) is recognizable as a Mandelbrot fractal, suitable for quantifying second-order conformational changes [5]. It arises because the longer segments self-similarly fold back on themselves, occluding the SASA of the central aa. The most surprising aspect of this folded occlusion is that its self-similarity is nearly universal on average, and almost independent of the individual protein fold.

The sculpting effects of billions of years of protein aqueous evolution have smoothed globular differential geometries, as described by (1) for all proteins. For a specific protein self-similar smoothing effects also occur, but now on a modular wave length determined by the protein's function. Given Ψ (aa), we calculate the modular average

$$\Psi \ (aa, W) \ = average(\Psi \ (aa - M), \Psi \ (aa + M) ) \qquad (2)$$

which is a rectangular window of width W = 2M + 1. We will look at minimal values of W for protein strains used in the HPV vaccine.

Before we do so, one more concept is needed. Level sets were developed to track the motions of continuum interfaces [6] – applied here to protein globular surfaces. Practical applications of level sets have emphasized image analysis [7], and have gradually evolved to include Voronoi partitioning, just as has been used for deriving protein hydropathicity scales since 1978 [8]. We expect, of course, that hydrophobic pivots move most slowly, while hydrophilic hinges move fastest. When there are two or more degenerate (level) pivots or hinges, it is likely that this is not accidental (nothing in proteins is), and we can test this assumption by comparing profiles with different scales. Synchronized motions should enable self-assembly [9].

## Results

The long road that led to cervical cancer vaccines began in 1976 when Harald zur Hausen published the Nobel hypothesis that human papilloma virus (HPV) plays an important role in the cause of cervical cancer. HPV is a large capsid protein, but it was found that only the L1 part was needed to make a good vaccine that conformationally self-assembled into morphologically correct virus-like particles (VLPs) [10]. L1 from HPV 16, taken from lesions that had not progressed to cancer, self-assembled $10^3$ times more efficiently than the HPV 16 L1P that researchers everywhere had been using; the old strain L1P had been isolated from a cancer, which differed from L1 by only a single amino acid mutation D202H [11]. Vaccines based on the unaffected strain L1 are also surprisingly effective even for many strains mutated at sites **S** = {76,176,181,191,282,353,389,474}, singly or in combinations of up to 6 mutations [12]. The dramatic effect of the single amino acid mutation D202H on HPV vaccine effectiveness is a long-standing mystery, which profiling with fractal $\Psi$ (aa,W) appears to solve.

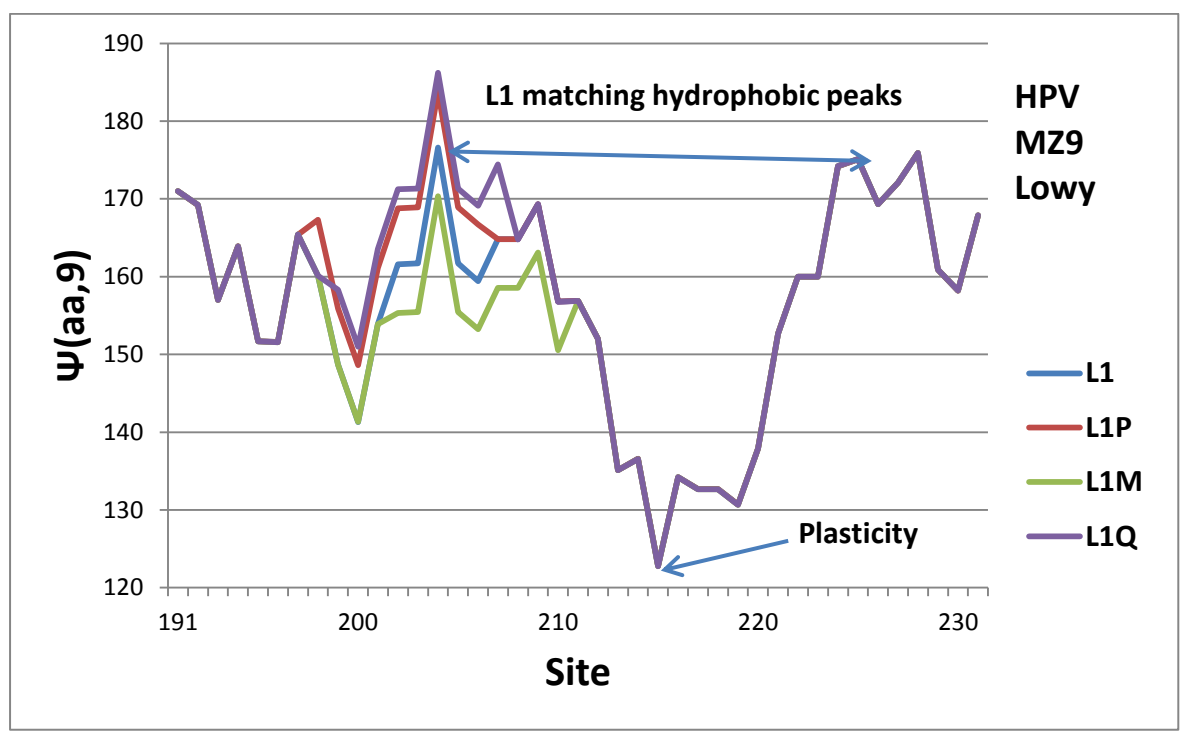

Fig. 1. Hydroprofile of L1 and several single amino acid mutants: L1P (D202H), L1Q (T203I), and L1M(G206S), using the fractal scale [4].

Given the lower bound of L = 9, and the remarkable properties of L1, its profile $\Psi$ (aa,9) with the fractal scale (1) was derived. The striking feature is the presence of two level L1 hydrophobic peaks between 191 and 231, shown enlarged in Fig.1. The narrow peak $\alpha$ is centered on 202, the mutated site distinguishing L1 from L1P. With W = 9, it spans the region 198-206. The triple peaks $\beta$ span the region 219-230. Note that none of the sites **S** overlap either the single peak region $\alpha$, or the triple peak region $\gamma$. Using BLAST, one can search NCBI lists of singly mutated 505 amino acid L1 proteins. If the mutations were random, among a sample of 150 such recent strains, one would expect to find about 2$\alpha$ and 3$\gamma$ mutations, but the list actually contains 5α and 0β mutations. The narrow peak mutates more often than random, while the broad peak appears to be very stable, in accord with the general properties of L1 proteins [10-13].

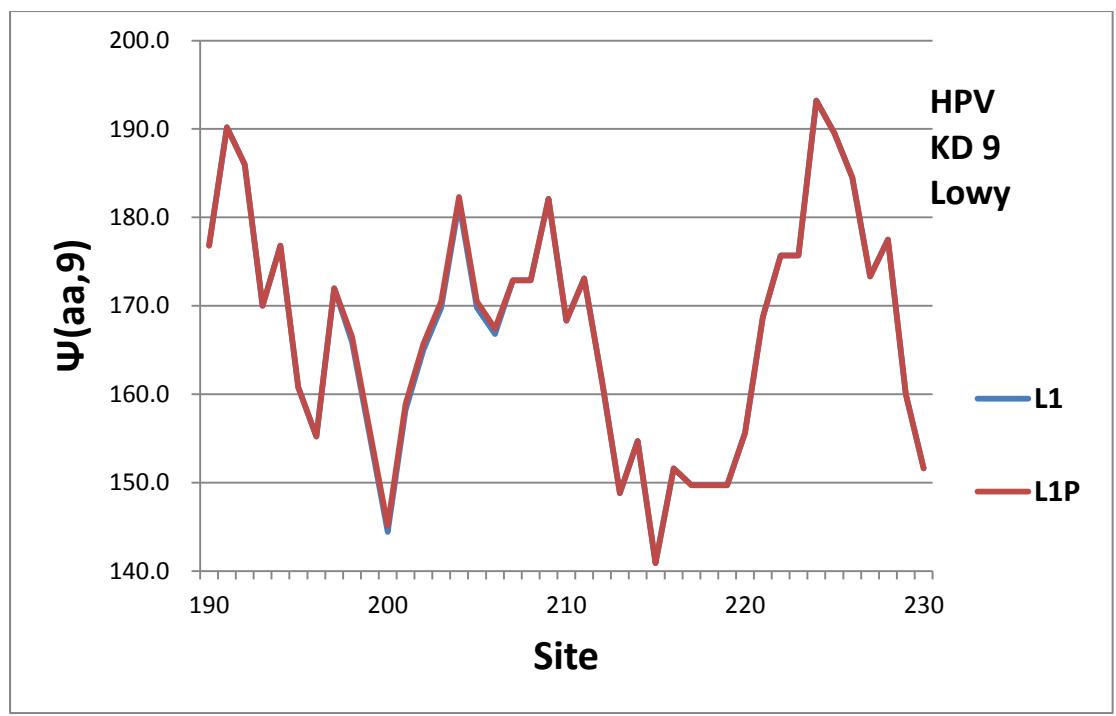

Fig. 2. Hydroprofiles of the critical region of L1 and L1P, using the standard thermodynamically first-order KD scale [13]. The small differences between L1 and L1P are resolvable at high resolution online.

## Discussion

The leveling in Fig. 1 can be used to test the accuracy of the fractal range [4]. Compared to the overall width of L1 $\Psi$ (aa,9), L1 peaks are level to 1%, while L1P, L1M and L1Q are level to only 5%. If we shift W up to 11 (in the fractal range), the results are similarly good, but if we decrease W to 7 (outside the fractal range), the L1 leveling is accurate to only 5%.

L1 peak leveling can also be used to test any protein hydropathicity scale, for instance, the standard 1982 unfolding scale "KD" based on enthalpy differences between air and water [14]. Results for the region shown in Fig. 1 for the fractal scale [3] are shown in Fig. 2. Overall the features are similar, but α,β peak leveling is no longer apparent. On the KD scale $\Psi$(D) nearly equals Ψ(H), so the difference between the two curves near 202 is visible only with high resolution; it is much smaller that the α,β difference.

## Conclusions

Most biomedical differences are too small to be resolved with any Ψ scale other than the best bioinformatic scale. These differences often involve allosteric (noncontact, or long-range) interactions, which fractal scaling has quantified for classic cases, such as aspirin [15], globins (especially neuroglobin, essential to neural high metabolic activity) [16], and amyloid aggregation [17]. Bioinformatic scaling is also supportive of emergent microarray epitopic blood profiling for early cancer detection [18-20].

**Methods** The calculations described here are very simple, and are most easily done on an EXCEL macro. The one used in this paper was built by Niels Voorhoeve and refined by Douglass C. Allan.

cells are critically self-organized systems synchronized by mechanical interactions. Proc. Nat. Acad. Sci. (USA) **108**, 13978-13983.

.